amsmath,amssymb]{revtex4}
\documentclass[prd,aps,preprintnumbers,%
amsmath,amssymb]{revtex4}

\usepackage{graphicx}   

\def\Pslash{\hbox{/\kern-.65em$P$}}
\def\Kslash{\hbox{/\kern-.65em$K$}}

\begin{document}

\title{Constraints on flavour-dependent long-range  forces from 
atmospheric
neutrino observations at Super-Kamiokande }

\author{Anjan S. Joshipura$^a$ and   Subhendra Mohanty $^{a,b}$}
\affiliation{$^a$ Physical Research Laboratory, Ahmedabad 380009,
India.\\ 
$^b$ International Center for Theoretical Physics, Trieste, Italy.}

\date{\today }

\begin{abstract}
In the minimal standard model it is possible to gauge any one of 
the following global symmetries in an anomaly free way: (i)
$L_{e}-L_{\mu}$,  (ii) $L_{e}-L_{\tau}$ or (iii)
$L_{\mu}-L_{\tau}$. If the gauge boson corresponding to (i) or (ii) is
(nearly) massless
then it will show up as a long range
composition dependent fifth force between macroscopic objects. 
Such a force will also influence neutrino oscillations due to its
flavour-dependence. We show that the latter effect is quite significant in
spite of very strong constraints on the relevant gauge couplings from the
fifth force
experiments. In particular,
the $L_{e}- L_{\mu,\tau}$ potential of the electrons in
the Sun and the earth is shown to suppress the atmospheric neutrino
$\nu_\mu \rightarrow \nu_\tau$ oscillations which have been
observed at Super-Kamiokande. The Super-K data of
oscillation of multi-GeV atmospheric neutrinos can be used to put an
upper bound
on coupling $\alpha_{e\tau}< 6.4 \times 10^{-52}$ and
$\alpha_{e\mu}< 5.5 \times 10^{-52}$at $90\%$ CL when the range
of the force is the earth-sun distance. This is an improvement by two
orders on the earlier fifth force bounds in this range.
 \end{abstract}

\maketitle

\begin{figure}
 \label{Fig.1}
\centering
\includegraphics[width=7cm]{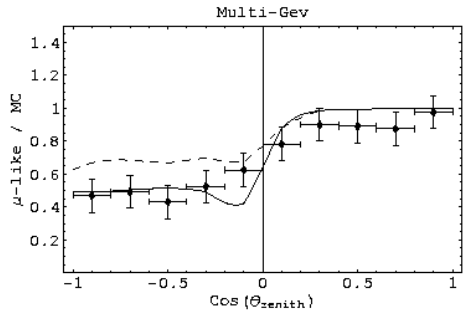}
\caption{ Observed Mu-like events/MC  in the multi-gev energy. The
solid line is a fit with $\alpha_{e \mu}=0$, $\Delta m_{23}^2=3.9
\times 10^{-3}$ ,$Sin^2 2 \theta_{23}=1$. The dashed curve is for
$\alpha_{e \mu}=5.5 \times 10^{-52}$ with the same values for  $\Delta
m_{23}^2$  and $Sin^2 2 \theta_{23}$}
\end{figure}

\begin{figure}
 \label{Fig.2}
\centering
\includegraphics[width=7cm]{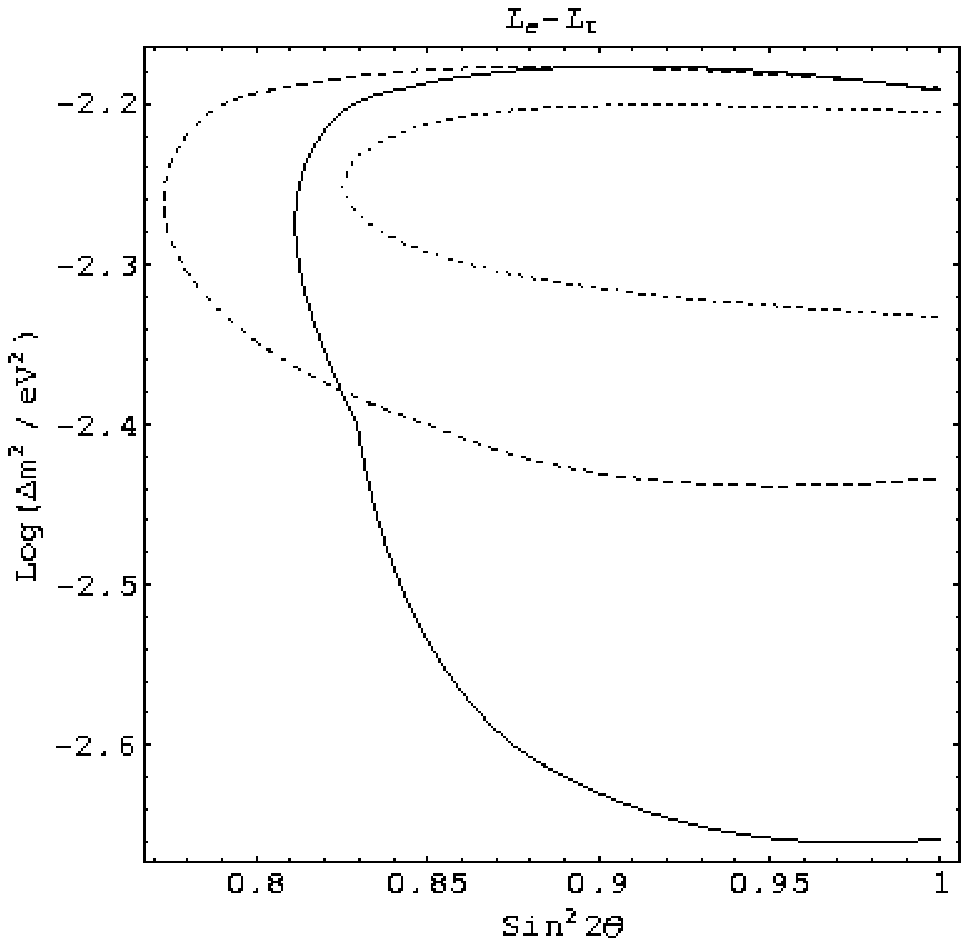}
\caption{\it $L_{e}-L_{\tau}$ gauge symmetry: Allowed values of
$\Delta m_{23}^2$ and $Sin^2 2 \theta_{23}$ at $90 C.L. \%$ with
$\alpha_{e \tau}=0 $(solid curve), $\alpha_{e\tau}=4\times
10^{-52} $ (dashed curve) and $\alpha_{e\tau}=5\times 10^{-52} $
(dotted curve). For $\alpha_{e\tau}=6.4\times 10^{-52} $ there is
no allowed parameter space  of $\Delta m_{23}^2$ and $Sin^2 2
\theta_{23}$ which is consistent with the Super-K atmospheric
neutrino data. }
\end{figure}

\begin{figure}
 \label{Fig.3}
\centering
\includegraphics[width=7cm]{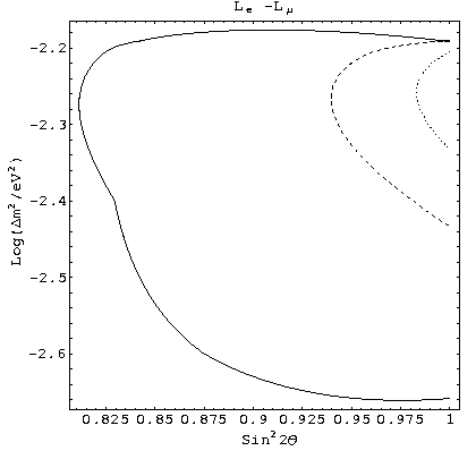}
\caption{ $L_{e}-L_{\mu}$ gauge symmetry: Allowed values of
$\Delta m_{23}^2$ and $Sin^2 2 \theta_{23}$ at $90 C.L. \%$ with
$\alpha_{e\tau}=0 $(solid curve), $\alpha_{e\mu}=4\times 10^{-52}
$ (dashed curve) and $\alpha_{e \mu}=5\times 10^{-52} $ (dotted
curve). For $\alpha_{e\mu}=5.5\times 10^{-52} $ there is no
allowed parameter space  of $\Delta m_{23}^2$ and $Sin^2 2
\theta_{23}$ which is consistent with the Super-K atmospheric
neutrino data. }
\end{figure}

The standard model is invariant under four global
symmetries corresponding to the baryon number and lepton numbers 
$L_\alpha$ of the three families ($\alpha=e,\mu,\tau$)
of leptons. None of these symmetry by themselves can be gauge
symmetries but there exists three
combinations each of which can be gauged in an anomaly free way
alongwith the standard $SU(2)\times U(1)$ group. These correspond
\cite{foot} to (i) $L_{e}-L_{\mu}$,  (ii) $L_{e}-L_{\tau}$ or (iii)
$L_{\mu}-L_{\tau}$. Recent experimental indications of neutrino
oscillations lead us to conclude that none of these three
symmetries can be an exact symmetry of nature since exact conservation of
the corresponding charges prevent  mixing of different neutrino species
and hence oscillations among them contrary to strong indications from the
solar, KamLand and atmospheric neutrino experiments.
Hence these symmetries must be broken in
nature. The phenomenological consequences of relatively heavy
gauge bsosons corresponding to these symmetries  have been discussed in
\cite{foot, vg}. Here we concentrate on
an alternative possibility corresponding to very light gauge bosons
with typical masses corresponding to a range greater than or equal to the
earth-sun distance. Such a scenario is strongly constrained by the  fifth 
force
experiments but it still has interesting consequences in neutrino physics
which we discuss.

For very light masses, the exchange of an $L_{e}-
L_{\mu,\tau}$
gauge boson between electrons will give rise to a composition
dependent long range force between macroscopic bodies.
A  variety of experiments have been performed to look for such
equivalence principle violating long range forces
\cite{adelberger}. A ($L_{e}- L_{\mu,\tau}$) gauge boson exchange
between electrons will give rise to a potential $V(r)= \frac{g^2}{4 \pi
r} e^{-r m_b}$ where $m_b$, the gauge boson mass determines the
range $\lambda$, of the force $\lambda =1/m_b$. For  composition
dependent  forces in the earth-sun distance range the most
stringent bounds come from lunar laser ranging  experiments
\cite{llr,llr2} and from earth based torsion balance experiments
where a search is made of a composition dependent force on torsion
balance which would be in phase which the diurnal rotation of the
earth \cite{heckel}. The earth and the moon have different $Z$
composition and in the presence of a solar distance $L_{e}-
L_{\mu,\tau}$ potential caused by the electrons in the sun, they
will have different acceleration towards the sun. From the
differential acceleration of the earth-moon system towards the sun
and torsion balance experiments one can put an upper bound of
$\alpha_{(e~\mu,\tau)} < 3.3 \times 10^{-50}$ for a $Z$ dependent
force with the range $\lambda \simeq 10^{13} cm$
\cite{adelberger}.

The basic observation of this paper is that in spite of very stringent
constraints on $\alpha_{e~\mu,\tau}$, the $L_e-L_{\mu,\tau}$ forces
can significantly influence the neutrino oscillations. This comes about
due to the long range nature and the flavour dependence of the
potential generated by the gauge bosons of the $L_e-L_{\mu,\tau}$
symmetry through the $\nu_{\mu,\tau}-e$ elastic scattering. For example,
the electrons inside the Sun generate a potential $V_{e~\mu,\tau}$ at the
earth surface which is given by
\begin{eqnarray}\label{vetau}
  V_{e ~\mu,\tau}= \alpha_{e ~\mu,\tau} \frac {N_e}{R_{es}}  
   = \alpha_{e~\mu, \tau}~ \frac{Y_e M_\odot}{
  m_n}\frac{1}{A.U.}= 3.3\cdot 10^{-11} eV \left(\frac{\alpha_{e~
\mu,\tau}}{3.3
\cdot 10^{-50}}\right)
\end{eqnarray}
where $\alpha_{e~\mu,\tau} \equiv g^2_{\mu,\tau}/4 \pi$ and $g_{\mu,\tau}$
is the
gauge coupling of the $L_e-L_{\mu,\tau}$ symmetry. The electron fraction
in the sun
$Y_e \sim
(2/3)$ , the solar mass $M_\odot =1.12 \times 10^{57}$ GeV, the earth sun
distance $A.U= 1.5 \times 10^{13} cm =7.6 \times 10^{26} {\rm GeV}^{-1}$
and the nucleon mass $m_n \simeq 0.939 GeV$ lead to the numerical value
quoted above.  
The corresponding potential due to electrons in the earth of an 
earth-radius range force is
about 20 times smaller. This means that the bounds on the $\alpha_{e 
\mu,\tau}$ established for solar-distance forces reduce by a factor
20 for earth distance forces i.e  $\alpha_{e \mu} < 1.1 \times 10^{-50}$ 
and  $\alpha_{e \tau} < 1.2 \times 10^{-50}$ for $\lambda \sim 6400 km $.
The improvement in the bound on earth-radius range fifth force is an 
improvement on the existing bounds \cite{adelberger} by more than five 
orders of magnitude.

The potential given in eq.(\ref{vetau}) is comparable or greater than 
the
$\frac{\Delta m^2}{E}$ probed in various neutrino experiments, e.g.
$\frac{\Delta m^2}{E}\sim (10^{-12}-10^{-14}) $eV in case of the
multi-GeV
atmospheric neutrinos. Thus the $V_{e \mu,\tau}$ can lead to
observable changes in the oscillations of the terrestrial, solar and
atmospheric neutrinos. At the very least, these experiments can be used to
put more stringent constraints on $\alpha_{e~\mu,\tau}$ than the
existing  fifth force experiments. We illustrate this through a study of
the atmospheric neutrinos.

The
observations of atmospheric neutrinos at Super-Kamiokande have
shown that the $\nu_e$  and $\bar \nu_e$ produced by cosmic rays
in the atmosphere are largely unaffected, whereas the $\nu_\mu$
and $\bar \nu_\mu$ get converted partially to $\nu_\tau$ and $\bar
\nu_\tau$ respectively \cite{Super-K}. The neutrino parameters
from the Super-K data are $\Delta m_{23}^2 = 2.8 \times 10^{-3}
eV^2$ and $Sin^2 2 \theta_{23} =1$ for vacuum oscillations. 
These parameters would get affected in the presence of the potential
(\ref{vetau}). 
For a small value of $\alpha_{(e~\mu,\tau)} (< 10^{-52})$ 
the oscillation probability
observed in Super-K can be reproduced by shifting $\Delta m^2$ and
$Sin^2 2 \theta_{23}$. With increase in
$\alpha_{e~\mu,\tau}$ the allowed parameter space becomes 
smaller and finally at some 
$\alpha_{e~\mu,\tau}$ no values of the parameters $\Delta_{23},
Sin^2 2\theta_{13}$ can  fit the Super-K multi-GeV muon neutrino event
data
\cite{data}. This way we obtain  a 90\% CL upper bound of $\alpha_{e \mu}
<
5.5 \times 10^{-52}$ and $\alpha_{e \tau} < 6.4 \times 10^{-52}$ on the
couplings  of the possible $L_e-L_\mu$
and $L_e -L_\tau$ gauge forces respectively.

{\it $L_{e}-L_{\tau}$ gauge symmetry :}
 The $\nu_{\mu} -\nu_{\tau}$ oscillations are governed by the
evolution equation
\begin{equation}\label{H}
 i \frac{d}{dt}  \begin{pmatrix}
 \nu_\mu  \\
   \nu_\tau \
  \end{pmatrix}=  \begin{pmatrix}
    -\frac{\Delta m_{23}^2}{4 E_\nu} Cos 2 \theta_{23} & \frac{\Delta m_{23}^2}{4 E_\nu} Sin 2 \theta_{23}  \\
    \frac{\Delta m_{23}^2}{4 E_\nu} Sin 2 \theta_{23}  & \frac{\Delta m_{23}^2}{4 E_\nu} Cos 2 \theta_{23}  - V_{e \tau} \
  \end{pmatrix}\begin{pmatrix}
    \nu_\mu  \\
    \nu_\tau \
  \end{pmatrix}
\end{equation}

Where the potential $V_{e \tau}$ is due to exchange of
$L_{e}-L_{\tau}$ gauge boson. For anti-neutrinos the potential in
(\ref{H}) will appear with a negative sign. The survival
probability of the atmospheric muon neutrinos can be written as
\begin{equation}\label{p}
  P_{\mu \mu}= 1- Sin^2 2\tilde \theta_{23} ~Sin^2 \frac{\Delta \tilde m_{23}^2
  L}{4 E_\nu}
\end{equation}
where the effective mixing angle $\tilde \theta_{23}$ and mass
squared difference $\Delta \tilde m_{23}^2$ are given in terms of
the corresponding vacuum quantities appearing in the Hamiltonian
(\ref{H}) by the relations
\begin{equation}\label{dM}
  \Delta \tilde m_{23}^2 =\Delta m_{23}^2 \left[(\xi_{e \tau}-Cos 2
\theta_{23})^2 +
  Sin^2  2\theta_{23})^{1/2}\right]
\end{equation}
and
\begin{equation}\label{sin}
  Sin^2 2 \tilde \theta_{23} = \frac{Sin^2 2 \theta_{23}}
{\left[(\xi_{e \tau}-Cos 2 \theta_{23})^2 +
  Sin^2 2 \theta_{23}\right]}
\end{equation}
where the strength of the potential is characterized by the
parameter
\begin{equation}\label{xi}
  \xi_{e \tau} \equiv  \frac{2 V_{e \tau} E_{\nu}}{\Delta m^2}
\end{equation}
with $V_{e\tau}$ given as in eq.(\ref{vetau}).
The $\bar \nu_\mu$ survival probability is obtained from the
$\nu_\mu$ survival probability by replacing $\xi \rightarrow -
\xi$ in the expressions (\ref{dM}) and (\ref{sin}).  

Note that there is a possibility of the resonant enhancement of the
atmospheric neutrino mixing angle  due to the MSW 
effect generated by $V_{e\tau}$. This depends on the sign of $\xi_{e\tau}$
and could occur either for neutrino or anti neutrino. Since the
atmospheric flux contains comparable fractions of both, this effect
would get washed out and one still needs large mixing angle to explain the
atmospheric data as our detailed analysis presented below shows.

In Super-Kamiokande the neutrino flavor is identified by the charged
current interaction $\nu_l + N \rightarrow N^\prime + l + X $
($l=e, \mu$). The outgoing muon or electron is identified by its
characteristic Cerenkov cone. We use multi-GeV and partially
contained  mu-like events data for 3 years of operation
\cite{data}. The ratio of the observed mu-like events to 
the corresponding Monte Carlo data 
\cite{data} in the multi-GeV range ($ E_\nu \sim (1-100) GeV$ ) is
shown as a function of the zenith-angle in Fig 1. The cosine of the 
zenith
angle, $Cos \theta_z$ is related to the neutrino flight path-length
$L$ in (\ref{p}) as
\begin{equation}\label{L}
  L= ((R_e + h)^2 - R_e
Sin^2 \theta_z)^{1/2} -R_e Cos \theta_z
\end{equation}
 where $R_e= 6374 km$ is
the mean radius of the earth and $h \simeq 15 km$ is the average
height in the atmosphere where the neutrinos are produced. 

The multi-GeV data are presented as bins in energy and cosine of the
zenith angle. We average over the energy bins in our analysis. The
theoretical prediction for the mu-like events/Monte Carlo can then be
written  as
\begin{equation}\label{mulike}
 \frac{(\mu-like)}{MC}(Cos \theta_z) =\frac{ \int dE_\nu ( P_{\mu
\mu}\Phi_\mu + r  P_{\bar
 \mu \bar \mu} \Phi_{\bar \mu})}{  \int dE_\nu ( \Phi_\mu + r  \Phi_{\bar \mu})  }
\end{equation}
where $P_{\mu \mu}$ and  $P_{\bar
 \mu \bar \mu}$ are the survival probabilities
 and $\Phi_\mu (E_\nu, Cos \theta_z)$ and 
$\Phi_{\bar \mu} (E_\nu, Cos \theta_z)$
are fluxes of the atmospheric $\nu_\mu$ and $\bar \nu_\mu$
respectively. We use the Fluka-3D
flux given in \cite{fluka} in our analysis. In (\ref{mulike}) $r$
is the ratio of the cross sections $\sigma_{\bar \nu
N}/\sigma_{\nu N}$ which is $\sim 0.5$ \cite{rmp}.  In writing 
eq.(\ref{mulike}), we have neglected small
energy dependent of the relevant charged current cross section given in
\cite{rmp}.

We calculate the chi-square for the $10$ zenith angle bins with
$Cos \theta_z =(-0.9 - 0.9 )$ as a function of the parameters
$\Delta m_{23}^2, Sin^2 2 \theta_{23}$ for different values of
$\alpha_{e \tau}$. For $\alpha_{e \tau}=0$, we find that the
minimum chi-square is $4.82$ and corresponds to the best fit values
 $\Delta m_{23}^2 =3.9 \times 10^{-3}
eV^2, Sin^2 2 \theta_{23}=1$. In Fig 2. we show the $90 \% C.L$
allowed region when the long-range force is taken to be zero
(solid line).
 We increase $\alpha_{e \tau}$ in small steps and observe that the
 $\chi^2_{min}$ increases from the $\alpha=0$ value, and the
 allowed parameter space of $\Delta m_{23}^2, Sin^2 2 \theta_{23}$
 shrinks as shown in Fig 2. We find that when $\alpha_{e \tau}
 =6.4 \cdot 10^{-52}$ there is no allowed parameter space which is
 consistent with the Super-K observations of muon-neutrino events
 in the multi-gev energy range. From this we derive the
 upper bound on $\alpha_{e \tau} < 6.4 \times 10^{-52}$ at $90 \%
 C.L$

{\it $L_e -L_\mu$ gauge symmetry } 
The $\nu_\mu
\rightarrow \nu_\tau$ oscillations are governed by the 
following evolution equation when the long range potential arise from the
exchange of the $L_e -L_\mu$ gauge bosons.
\begin{equation}\label{H2}
i \frac{d}{dt} \begin{pmatrix}
    \nu_\mu  \\
    \nu_\tau \
  \end{pmatrix}=  \begin{pmatrix}
    -\frac{\Delta m_{23}^2}{4 E_\nu} Cos 2 \theta_{23}-  V_{e \mu}& \frac{\Delta m_{23}^2}{4 E_\nu} Sin 2 \theta_{23}  \\
    \frac{\Delta m_{23}^2}{4 E_\nu} Sin 2 \theta_{23}  & \frac{\Delta m_{23}^2}{4 E_\nu} Cos 2 \theta_{23}  \
  \end{pmatrix}\begin{pmatrix}
    \nu_\mu  \\
    \nu_\tau \
  \end{pmatrix}
\end{equation}

The expression for the $\nu_\mu$ and the $\bar \nu_\mu$ survival
probabilities are identical to (\ref{p},\ref{dM},\ref{sin}) with
$V_{e\tau}$
replaced with $-V_{e\mu}$. Therefore, the survival probabilities of
$\nu_\mu$ and $\bar \nu_\mu$ in case of the 
$L_e -L_\tau$ and the $L_e-L_\mu$ symmetry satisfy the following
relations:
\begin{equation}\label{ppba}
  P_{\mu \mu}(V_{e \tau})= P_{\bar \mu \bar \mu}(-V_{e \tau})=P_{\mu \mu}(-V_{e
  \mu}) = P_{\bar \mu \bar \mu}(V_{e \mu})
\end{equation}

Using the same procedure as discussed
above we find that in case of the $L_e -L_\tau$ symmetry, the upper bound
on the coupling constant is
$\alpha_{e \mu} < 5.5 \times 10^{-52}$  (Fig 3).

We have concentrated here on the $\nu_\mu-\nu_\tau$
oscillations and $\mu$-like events at Super-K. In general, the 
$\nu_e-\nu_{\mu}$ oscillations would also get affected by the presence 
of the additional potentials considered here. When the $\nu_e-\nu_\mu$ 
oscillations are governed by the solar scale, the survival probability of
the atmospheric electron neutrinos is nearly one. Addition of potential
tend to only suppress the $\nu_e-\nu_\mu$ oscillations  and one would not get any bound from the
study of the electron-like events at Super-K. There
would exist some
limited ranges of parameters where these oscillations would be resonantly
enhanced due to the contribution from $V_{e\mu,\tau}$. Such parameter
space would any way be ruled out from the non-observations of the
atmospheric electron neutrinos.

While we concentrated on the atmospheric neutrinos, $\Delta m_{12}^2/2 E$
in case of the solar neutrinos is not 
significantly larger than the value of the potential in eq.(\ref{vetau})
and the parameter $\xi$ in eq.(\ref{xi}) can become comparable to $\cos 2
\theta_{solar}$ for
the upper bound on $\alpha_{e~\mu,\tau}$ found here . This would effect
the effective
solar mixing angle both inside the Sun and at the earth. Thus
the $L_{e,\mu,\tau}$ would be expected to produce observable effects
on the solar neutrino oscillations also.

Let us now give a possible example of the theoretical generation
of the oscillation parameters in the presence of the $L_e-L_{\mu,\tau}$
symmetry. We choose a specific case of the $L_e-L_\tau$ symmetry. Without
specifying underlying mechanism, we assume that neutrino
masses  are generated by effective five dimensional operators constructed
from the standard model fields and an additional Higgs doublet $\phi'$
having the $L_e-L_\tau$ charge -1. The gauge invariance of the model
implies the following structure:

\begin{eqnarray} \label{five}
-L_m&=&\frac{1}{2}\frac{\phi^T\tau_2\tau^a \phi}{M}
\left[\beta_{13}(l^T_e C \tau_2\tau^a 
l_\tau+\beta_{22}l_\mu^TC \tau_2\tau^a l_\mu\right] \nonumber \\
&+&\frac{1}{2}\left[\frac{\phi^T\tau_2\tau^a \phi'}{M} \beta_{12}(l^T_e 
C \tau_2\tau^a
l_\mu)+
\frac{\phi^{'T}\tau_2\tau^a\phi'}{M} \beta_{11}(l^T_e C \tau_2\tau^a l_e)
\right]~,  
\end{eqnarray} 
where $M$ is a high scale associated with the physics generating the above
operators, e.g. scale of the right handed neutrinos in the seesaw model.
We have suppressed Lorentz indices in writing above equation. $C$ refers
to the usual charge conjugate matrix, $l$ to the leptonic doublet and
$\tau_2,\tau_a (a=1,2,3)$ act in
the $SU(2)$ space.

Eq.(\ref{five}) generates the following neutrino mass matrix  
\begin{equation} \label{mnu}
M_\nu=\left(\begin{array}{ccc}
m_{11}&m_{12}&m_{13}\\
m_{12}&m_{22}&0\\
m_{13}&0&0\\ \end{array} \right) ~.
\end{equation}

The $m_{ij}$ are proportional to $\beta_{ij}$ and can be read off from
eq.(\ref{five}).
In the exact symmetry limit corresponding to $<\phi'>=0$
the above mass matrix describes a Dirac neutrino with mass $m_{13}$
and a
Majorana neutrino with mass $m_{22}$. These remain unmixed and there are
no
neutrino oscillations. A non-zero $<\phi'>$ leads to the required mixing
and mass splitting. If parameters $\beta$ and relevant vacuum
expectation values (vevs) in
eq.(\ref{five}) are chosen 
to give a hierarchy $m_{11}\leq m_{22}\ll m_{12}\sim m_{13}$ then the
above
mass matrix displays an approximate
$L_e-L_\mu-L_\tau$ symmetry. This symmetry is known \cite{emt} to lead to
the
successful explanation of the atmospheric neutrino problem. The presence
of $m_{11},m_{22}$ breaks this symmetry and generates the splitting
required to explain
the solar neutrino oscillations.

It is interesting to note that although we need very light gauge boson
$Z_\tau$ with typical mass $M_{Z_\tau}\leq 4.8\times 10^{-22}$ MeV
corresponding
to the radius
of the earth, we do not need to fine tune the symmetry breaking vev
$<\phi'>$ to such an extent. This follows since $M_{Z_\tau}\sim g_\tau
<\phi'>$ and since $g_\tau<8.9~\times 10^{-26}$, a value of $<\phi'>$ in
few
GeV range would still
keep $M_{Z_\tau}$ very light. Likewise, the $Z-Z_\tau$ mixing
$\theta_{Z-Z_\tau}\sim \frac{g_\tau <\phi'>}{g <\phi>}$ also remains
very
small and does not lead to observable effects such as shift in the $Z$
mass. 

We have neglected possible Debye-screening of the long range forces
due to the solar plasma in
our
analysis above. This is justified since the screening length for the force
is given by $r_D= (4 \pi \alpha N_e /T)^{-1/2}$. In the center of the sun 
the electron density is $N_e\simeq 10^{26} cm^{-3}$ and the temperature is
$T=1.5\times 10^7 K$. Even for $\alpha $ as large as $10^{-49}$ the Debye
screening length turns out to be $r_D \simeq 10^{16} cm \simeq 10^3 A.U$.
So in the range of $\alpha $ which we discuss there will be no screening
at earth-sun distance. The electron density in the sun drops off faster
than the temperature so the average $r_D$ of the sun is larger than that
in the core.

 We conclude that the atmospheric neutrino
observations
at Super-Kamiokande enable us to put bounds on  long 
range equivalence
principle violating forces which are  two orders of magnitude 
more
stringent than the corresponding bounds from the classic fifth
force experiments.


\begin{thebibliography}{99}
\bibitem{foot} R.Foot, Mod. Phys. Lett.  {\bf A 6}, 527 (1991);
X.-G. He, G.C.Joshi, H. Lew and R.R. Volkas, Phys. Rev {D 44}, 2118
(1991); R. Foot {\it et al} Phys. Rev. {\bf D50} 4571 (1994).
\bibitem{vg} G. Dutta, A.S.Joshipura and K. B. Vijayakumar, Phys. Rev 
{D 50}, 2109 (1994).
\bibitem{adelberger}E.Fischbach and C.L.Talmadge, {\it The search
for non-Newtonian gravity}, New York, Springer Verlag (1999); E.G.
Adelberger, B.R. Heckel, A.E. Nelson,
 e-Print Archive:hep-ph/0307284

\bibitem{llr}
J.~G.~Williams, X.~X.~Newhall and J.~O.~Dickey,
Phys.\ Rev.\ D {\bf 53}, 6730 (1996).
\bibitem{llr2} J. Muller et al., In "{\it Proc of 8th Marcel
Grossman meeting on General Relativity}, Jerusalem (1997).

\bibitem{heckel} B.R.Heckel et al, In {\it Advances in Space
Research, Proc. of 32nd COSPAR Scient.Assembly}, Nagoya (1998).

\bibitem{Super-K} Y. Fukuda et al. Phys Rev Lett. {\bf 82}, 2644
(1999); Y. Fukuda et al. Phys Lett . {\bf B388}, 397 (1999);Y.
Fukuda et al. Phys Rev Lett. {\bf 85}, 3999 (2000).


\bibitem{data} J. Kameda in {\it Proc. Int. Cosmic Ray Conf.,}
Hamburg 2001,ed. by K.-H. Kampert et al, Vol 3, p 1162 (2001).

\bibitem{fluka} G. Battistom {\it et al}, hep-ph/207035.

\bibitem{rmp} T. Kajita and Y. Totsuka, Rev. Mod. Phys. {\bf 73} 2001 85.

\bibitem{emt} R. Barbieri {\it et al}, Phys. Lett. {\bf B445}
 239 (1999); A. S. Joshipura, Phys. Rev. {\bf D 60}  053002 (1999);
A. S. Joshipura and S. D. Rindani, Phys. Lett. {\bf B 464}  239 (1999) 
and Euro. Phys. Journal, {\bf C14}  85 (2000); H. S. Goh, R. N. 
Mohapatra and S. P. Ng, Phys. Lett.{\bf 542} 116 (2002); L. Lavoura,
Phys Rev. { \bf D 62}  093011 (2002); W. Grimus and L. Lavoura,
Phys. Rev. {\bf D62} 093012 (2000);Phys. Rev. {\bf D62} 093012
(2000); J. of High Energy Phys. {\bf
07} (2001) 045; R. N. Mohapatra, Phys. Rev. {\bf D64} 091301
(2001); K. S. Babu and
R. N. Mohapatra, Phys. Lett.{\bf B532} 77 (2002);
R. Kuchimanchi and R. N. Mohapatra,Phys. Rev. {\bf D66} 051301 (2002).   
%
\end{thebibliography}
\end{document}